\documentclass[twocolumn,prd,unsortedaddress,superscriptaddress,showpacs,a4paper,nofootinbib]{revtex4}
\usepackage{epsfig}

\def\beq{\begin{equation}}
\def\enq{\end{equation}}
\def\beqa{\begin{eqnarray}}
\def\enqa{\end{eqnarray}}

\def\Tr{\mbox{ Tr }}

\def\uu{\lag\bar{u}u\rag}
\def\dd{\lag\bar{d}d\rag}

\def\G3{\lag g^3G^3\rag}
\def\pli{p^\prime}

\def\la{\lambda}

\def\ga{\gamma}
\def\Ga{\Gamma}

\def\si{\sigma}

\def\al{\alpha}

\def\lb{\label}
\def\nnb{\nonumber}
\def\nn{\nonumber}

\newcommand{\rag}{\rangle}
\newcommand{\lag}{\langle}

\def\MeV{\mbox{ MeV}}
\def\GeV{\mbox{ GeV}}
\begin{document}

\title{\sc QCD Sum Rules for the X(3872) as a mixed molecule-charmoniun state}
\author{R.D.~Matheus}
\email{matheus@if.usp.br}
\affiliation{Instituto de F\'{\i}sica, Universidade de S\~{a}o Paulo, 
C.P. 66318, 05389-970 S\~{a}o Paulo, SP, Brazil}
\author{F.S.~Navarra}
\email{navarra@if.usp.br}
\affiliation{Instituto de F\'{\i}sica, Universidade de S\~{a}o Paulo, 
C.P. 66318, 05389-970 S\~{a}o Paulo, SP, Brazil}
\author{M.~Nielsen}
\email{mnielsen@if.usp.br}
\affiliation{Instituto de F\'{\i}sica, Universidade de S\~{a}o Paulo, 
C.P. 66318, 05389-970 S\~{a}o Paulo, SP, Brazil}
\author{C.M.~Zanetti}
\email{carina@if.usp.br}
\affiliation{Instituto de F\'{\i}sica, Universidade de S\~{a}o Paulo, 
C.P. 66318, 05389-970 S\~{a}o Paulo, SP, Brazil}

\begin{abstract}
We use QCD sum rules to test the nature of the meson $X(3872)$, assumed to be 
a mixture between charmonium and exotic molecular $[c\bar{q}][q\bar{c}]$
states with $J^{PC}=1^{++}$. We find that there is only a small range
for the values of the mixing angle, $\theta$, that can provide simultaneously 
good agreement with the experimental value of the mass and the decay width,
and this range is $5^0\leq\theta\leq13^0$. In this range we get
$m_X=(3.77\pm0.18)$  GeV and $\Gamma(X\to J/\psi\pi^+\pi^-)=(9.3\pm6.9)$ MeV,
which are compatible, within the errors, with the experimental values.
We, therefore, conclude that the $X(3872)$ is approximately 97\% a charmonium
state with 3\% admixture of $\sim$88\% $D^0D^{*0}$ molecule and $\sim$12\% $D^+
D^{*-}$ molecule. 
\end{abstract}

\maketitle

%
%
\section{Introduction}
%
%
Among the new hadronic states discovered in the last few years, the $X(3872)$ 
is one of the most interesting. It has been first observed by the 
Belle collaboration in the 
decay $B^+\!\rightarrow\!X(3872)K^+\rightarrow\!J/\psi\pi^+\pi^- K^+$
\cite{belle1}. This observation was later confirmed by CDF, D0 and 
BaBar \cite{Xexpts}.  The current world average mass is
$m_X=(3871.4\pm0.6)\MeV $ which is at the threshold for the production of the 
charmed meson pair  ${D^0}{\bar D}^{0\ast}$.
This state is extremely narrow, with a 
width smaller than 2.3 MeV at 90\% confiedence level.
Both Belle and Babar collaborations reported the radiative decay
mode $X(3872)\to \gamma J/\psi$ \cite{belleE,babar2}, which
determines $C=+$.  Further studies from Belle and CDF that combine
angular information and kinematic properties of the $\pi^+\pi^-$ pair,
strongly favor the quantum numbers $J^{PC}=1^{++}$ or $2^{-+}$ 
\cite{belleE,cdf2,cdf3}. 

In  constituent quark models \cite{bg} the masses of the possible charmonium
states with $J^{PC}=1^{++}$ quantum numbers are: $2~^3P_1(3990)$ and
$3~^3P_1(4290)$, which are much bigger than the observed mass. In view of this
large mass discrepancy the attempts to understand the $X$ meson as a 
conventional 
quark-antiquark states were abandoned. The next possibility explored was to 
treat 
this state as a multiquark state, composed by $c$, $\overline{c}$ 
and a light quark antiquark pair. Another experimental finding in favor of this 
conjecture is the the fact that the decay rates of the processes  
$X(3872) \to J/\psi\,\pi^+\pi^-\pi^0$ and 
$X(3872)\rightarrow\!J/\psi\pi^+\pi^-$ are comparable \cite{belleE}: 
\beq
{X \to J/\psi\,\pi^+\pi^-\pi^0\over X\to\!J/\psi\pi^+\pi^-}=1.0\pm0.4\pm0.3.
\label{rate}
\enq
This ratio indicates a  strong isospin and G parity violation, which is
incompatible with a $c\bar{c}$ structure for $X(3872)$. 

In a multiquark approach we can avoid the isospin violation problem. The 
next natural question is: is the $X$ made by four quarks in a bag or by a 
meson-meson molecule?

The observation of the above mentioned decays, plus the coincidence between 
the $X$ mass and the $D^{*0}D^0$ threshold: $M(D^{*0}D^0)=(3871.81\pm0.36)\MeV$
\cite{cleo}, inspired the proposal that the $X(3872)$ could be a molecular
$(D^{*0}\bar{D}^0-\bar{D}^{*0}D^0)$ bound state with small binding energy
\cite{close,swanson}. 
The $D^{*0}\bar{D}^0$ molecule is not an isospin eigenstate and the rate in 
Eq.(\ref{rate}) could be explained in a very natural way in this model.

Maiani and collaborators \cite{maiani} suggested that $X(3872)$ is a 
tetraquark. They have considered diquark-antidiquark states with $J^{PC}=
1^{++}$  and symmetric spin distribution:
\beq
X_q=[cq]_{S=1}[\bar{c}\bar{q}]_{S=0}+[cq]_{S=0}[\bar{c}\bar{q}]_{S=1}.
\enq
The isospin states with $I=0,~1$ are given by:
\beq
X(I=0)={X_u+X_d\over\sqrt{2}},\;\;\;\;X(I=1)={X_u-X_d\over\sqrt{2}}.
\enq
In \cite{maiani} the authors argue that the physical states 
are closer to mass eigenstates and are no longer isospin eigenstates. 
The most general states are then:
\beq
X_l=\cos{\theta}X_u+\sin{\theta}X_d,\;\;\;\;X_h=\cos{\theta}X_d-\sin{\theta}
X_u,
\enq
and both can decay into $2\pi$ and $3\pi$. Imposing the rate in 
Eq.(\ref{rate}), they obtain  $\theta\sim20^0$. They also argue that if $X_l$
dominates $B^+$ decays, then $X_h$ dominates the $B^0$ decays and 
vice-versa. Therefore, the $X$ particle in $B^+$ and $B^0$ decays would be  
different with  \cite{maiani,polosa} $M(X_h)-M(X_l)=(8\pm3)\MeV$. 
There are indeed  reports from Belle \cite{belleB0} and Babar  \cite{babarB0}
Collaborations on  the observation of the $B^0\to K^0~X$ decay. However, 
these reports (not completely consistent with each other) point to a
mass difference  much smaller than the predicited $ \simeq 8 \MeV$.  

All the conclusions in  ref.~\cite{maiani} were obtained in the context of a 
quark model. Given the uncertainties inherent to hadron spectroscopy, it is 
interesting to confront these theoretical results with QCD sum rules (QCDSR) 
calculations. 
This was partly done in  \cite{x3872} where, using the same tetraquark 
structure 
proposed  in ref.~\cite{maiani}, the  mass difference $M(X_h)-M(X_l)$ was 
computed  
and found to be in agreement with the   BaBar measurement 
($ M(X_h)-M(X_l)=(3.3\pm0.7)\MeV$). The same calculation \cite{x3872} has 
obtained
$m_X=(3.92\pm0.13)\GeV$. In QCDSR we can also use a current with the features 
of the mesonic molecule of the type $(D^{*0}\bar{D}^0-\bar{D}^{*0}D^0)$.  
With such a current the calculation reported in \cite{lnw} obtained the mass  
$m_X=(3.87\pm0.07)\GeV$ in a better agreement with the experimental mass. 
Therefore, from a QCDSR point of view, the $X(3872)$ seems to be better 
described with a  $D^*D$ molecular current than with a diquark-antidiquark 
current.  We feel though that 
the subject deserves further investigation.

In this work we use again the QCDSR approach to the $X$ structure including a 
new 
possibility: the mixing between two and four-quark states. This will be 
implemented  folowing the prescription suggested in \cite{oka24} for the light 
sector. The mixing 
is done at the level of the currents and will be extended to the charm sector. 
In a different context (not in QCDSR), a similar mixing was suggested already 
some time ago by Suzuki \cite{suzuki}. Physically, this corresponds to a 
fluctuation of the $c \overline{c}$ state where a gluon is emitted and 
subsequently splits into a
light quark-antiquark pair, which lives for some time and behaves like a 
molecule-like state. As it will be seen, in order to be consistent with  $X$ 
decay data, we must consider a second mixing  between:  
$(c \overline{c}) \, + \, (D^{*0}\bar{D}^0 -\bar{D}^{*0}D^0)$ and  
$(c \overline{c}) \, + \, (D^{*+}\bar{D}^- -\bar{D}^{*-}D^+)$.

With all these ingredients we perform a calculation of the mass of the $X(3872)$
and its decay width into $2 \pi$ and $3 \pi$.

%
%
\section{The mixed two-quark / four quark operator}
%
%

There are some experimental data on the $X(3872)$ meson that seem to
indicate the existence of a $c\bar{c}$ component in its structure. In 
ref.~\cite{suzuki} it was shown that, because of the very loose
binding of the molecule, the production rates of a pure $X(3872)$ molecule 
should be at least one order of magnitude smaller than what is seen
experimentally. Also, the recent observation, reported by BaBar \cite{babar09}, 
of the decay $X(3872)\to \psi(2S)\gamma$ at a rate:
\beq
{{\cal B}(X \to \psi(2S)\,\gamma)\over {\cal B}(X\to\psi\gamma)}=3.4\pm1.4,
\label{rategaexp}
\enq
is much bigger than the molecular prediction ~\cite{swan1}:
\beq
{\Gamma(X \to \psi(2S)\,\gamma)\over\Gamma(X\to\psi\gamma)}\sim4\times10^{-3}.
\label{ratega}
\enq
While this difference could be interpreted as a strong point against the
molecular model and as a point in favor of a conventional charmonium 
interpretation, it can also be  interpreted as an indication
that there is a significant  mixing of the $c\bar{c}$ component with 
the $D^0\bar{D}^{*0}$ molecule. Similar conclusion was also reached in
refs.~\cite{li1,li2}.
Therefore, we will follow ref.~\cite{oka24}
and consider a mixed charmonium-molecular current to study the $X(3872)$
in the QCD Sum Rule framework.

For the charmonium part we use the conventional axial current:
\beq
j'^{(2)}_{\mu}(x) = \bar{c}_a(x) \gamma_{\mu} \gamma_5 c_a(x).
\lb{curr2}
\enq

The $D^0$ $D^{*0}$ molecule is interpolated by \cite{liuliu,dong,stancu}:
\beqa
j^{(4u)}_{\mu}(x) & = & {1 \over \sqrt{2}}
\bigg[
\left(\bar{u}_a(x) \gamma_{5} c_a(x)
\bar{c}_b(x) \gamma_{\mu}  u_b(x)\right) \nonumber \\
& - &
\left(\bar{u}_a(x) \gamma_{\mu} c_a(x)
\bar{c}_b(x) \gamma_{5}  u_b(x)\right)
\bigg],
\lb{curr4}
\enqa
As in ref.~\cite{oka24} we define the normalized two-quark current as
\beq
j^{(2u)}_{\mu} = {1 \over 6 \sqrt{2}} \uu j'^{(2)}_{\mu},
\enq
and from these two currents we build the following  mixed charmonium-molecular 
current for the $X(3872)$:
\beq
J_{\mu}^u(x)= \sin(\theta) j^{(4u)}_{\mu}(x) + \cos(\theta) j^{(2u)}_{\mu}(x). 
\lb{field}
\enq

%
%
\section{The two point correlator}
%
%
%

%
The QCD sum rules \cite{svz,rry,SNB} are constructed from the two-point 
correlation function
\beqa
\Pi_{\mu\nu}(q) 
 = 
i\int d^4x ~e^{iq.x}\lag 0
|T[J_\mu^u(x)J_\nu^{u\dagger}(0)]
|0\rag = 
\nonumber \\  = 
-\Pi_1(q^2)\left(g_{\mu\nu}-{q_\mu q_\nu\over q^2}\right)+\Pi_0(q^2){q_\mu
q_\nu\over q^2}. 
\lb{2po}
\enqa
As the axial vector current is not conserved, the two functions,
$\Pi_1$ and $\Pi_0$, appearing in Eq.~(\ref{2po}) are independent and
have respectively the quantum numbers of the spin 1 and 0 mesons.

The sum rules approach is based on the principle
of duality.  It consists in the assumption that
the correlation function may be described at both
quark and hadron levels. At the hadronic level (the phenomenological side)
the correlation function is calculated introducing hadron characteristics
such as masses and coupling constants.  At the quark level,
the correlation function is written in in terms of
quark and gluon fields and a Wilson's
operator product expansion (OPE) is used to deal with
the complex structure of the QCD vacuum.

The phenomenological side is treated by first 
parametrizing the coupling of the axial vector meson
$1^{++}$, $X$, to the current, $J_\mu^u$, in Eq.~(\ref{field}) in terms
of the meson-current coupling parameter $\lambda^u$: 
\beq\label{eq: decay}
\lag 0 |
J_\mu^u|X\rag =\lambda^u\epsilon_\mu~. 
\label{lau}
\enq
Then, by inserting intermediate states for
the meson $X$, we can write the phenomenological side
of Eq.~(\ref{2po})  as 
\beq
\Pi_{\mu\nu}^{phen}(q)={(\lambda^u)^2\over
m_X^2-q^2}\left(-g_{\mu\nu}+ {q_\mu q_\nu\over m_X^2}\right)
+\cdots\;, \lb{phe} \enq
where the Lorentz structure projects out the $1^{++}$ state.  The dots
denote higher mass axial-vector resonances. This ressonances will be
dealt with through the introduction of a continuum
threshold parameter $s_0$.

In ref.~\cite{2hr} it was argued that a single pole ansatz can be problematic
in the case of a multiquark state, and that the two-hadron reducible (2HR)
contribution (or $S$-wave $D\bar{D}^*$ contribution, in the present case) 
should also be considered in the phenomenological side. However, in
ref.~\cite{2hr2} it was shown that the 2HR contribution is very small. The 
reason for this is the following. The 2HR contribution, in our case, can be 
written as \cite{2hr2}:
\beqa
\Pi_{\mu\nu}^{2HR}(q)&=&i(\lambda_{DD^*})^2\int~{d^4p\over(2\pi)^4}
\bigg({-g_{\mu\nu}+ p_\mu p_\nu/ m_{D^*}^2\over p^2-m_{D^*}^2}
\nn\\
&\times&{1\over(p-q)^2-m_D^2}\bigg), 
\lb{2hr} 
\enqa
where
\beq
\langle0|j_\mu^{(4u)}|DD^*(p)\rangle=\lambda_{DD^*}\varepsilon_\mu(p).
\enq
Following ref.~\cite{2hr2} the current two-meson coupling: $\lambda_{DD^*}$,
can be written in terms of the $D$ meson decay constant, $f_D$, and
the coupling of the $D^*$ meson with a 4-quark current. This last quantity 
should be very small, because the properties of the $D^*$ meson,  both in 
spectroscopy and in scattering, are very well understood if it is an ordinary 
quark-antiquark state. Therefore, the parameter $\lambda_{DD^*}$, should be 
very small, as in the case of the pentaquark \cite{2hr2}, and the 2HR 
contribution can be safely neglected.

In the OPE side we work up to dimension 8 at the leading order in $\alpha_s$. 
The
light quark propagators are calculated in coordinate-space and then Fourier 
transformed
to the momentum space. The charm quark part is calculated directly into the 
momentum space, with
finite $m_c$, and combined with the light part. The correlator in 
Eq.~(\ref{2po}) can be written as:
\beqa
\Pi_{\mu\nu}(q) & = &
\left({\uu \over 6 \sqrt{2}}\right)^2 \cos^2(\theta)
\,\Pi^{(2,2)}_{\mu\nu}(q) 
+ \nonumber \\ & + &
{\uu \over 6 \sqrt{2}}\left(\sin(2\theta)\right) 
\,\Pi^{(2,4)}_{\mu\nu}(q)
+ \nonumber \\ & + &
\sin^2(\theta)\, \Pi^{(4,4)}_{\mu\nu}(q),
\enqa  
with:
\beq
\Pi^{(i,j)}_{\mu\nu}(q) 
 = 
i\int d^4x ~e^{iq.x}\lag 0
|T[j^{(i)}_\mu(x)j^{(j)\dagger}_\nu(0)]
|0\rag.
\enq

 After
making a Borel transform of both sides, and
transferring the continuum contribution to the OPE side, the sum rule
for the axial vector meson 
up to dimension-eight condensates can
be written as:
\beqa (\lambda^u)^2e^{-m_X^2/M^2} 
 = \nonumber \\ =
\left({\uu \over 6 \sqrt{2}}\right)^2 \cos^2(\theta)
\,\Pi^{(2,2)}_{1}(M^2) 
 + \nonumber \\  + 
{\uu \over 6 \sqrt{2}}\left(\sin(2\theta)\right) 
\,\Pi^{(2,4)}_{1}(M^2)
+ \nonumber \\  +\, 
\sin^2(\theta)\, \Pi^{(4,4)}_{1}(M^2), 
\lb{sr} 
\enqa
where:
\beq
\Pi^{(2,2)}_{1}(M^2) 
= 
\!\!\int_{4m_c^2}^{s_0}\!\!\!\!ds~e^{-s/M^2}\rho_{pert}^{(22)}(s)  \;
+ 
\Pi^{(22)}_{\langle G^2\rangle}(M^2),\;
\enq
\beq
\Pi^{(2,4)}_{1}(M^2) 
= 
\!\!\int_{4m_c^2}^{s_0}\!\!\!\!ds~e^{-s/M^2}\rho^{(24)}_{\uu}(s)  \;
+ 
\Pi^{(24)}_{\langle\bar{u}G u\rangle}(M^2),\;
\enq
\beqa
\Pi^{(4,4)}_{1}(M^2) 
& = & 
\!\!\int_{4m_c^2}^{s_0}\!\!\!\!ds~e^{-s/M^2} 
\Bigg[
\rho_{pert}^{(44)}(s) + \rho^{(44)}_{\uu}(s)
+ \nonumber \\ & + &
\rho^{(44)}_{\uu^2}(s) + \rho^{(44)}_{\langle G^2\rangle}(s)
+ \rho^{(44)}_{\langle\bar{u}Gu\rangle}(s)
\Bigg]
+ \nonumber \\ & + &
\Pi^{(44)}_{\uu\langle\bar{u}Gu\rangle}(M^2),\;
\enqa
and
\begin{equation}
\rho_{pert}^{(22)}(s)=\frac{s}{4\pi^2}\left(1-
\frac{4m_c^2}{s}\right)^\frac{3}{2},\;
\label{OPE.first}
\end{equation}
\beqa
\Pi^{(22)}_{\langle G^2\rangle}(M^2)
& =  -\frac{\langle g^2G^2\rangle}{3\cdot 2^5\pi^2}
\int_0^1d\alpha 
\Bigg[
\frac{2\alpha(1-\alpha)M^2+m_c^2}{\alpha(1-\alpha)M^2}
+ \nonumber \\ &  + 
2 m_c^2 
\frac{(2\alpha-1)m_c^2+\alpha(\alpha^2-1)M^2}{M^4\alpha^3(\alpha-1)}
\Bigg] e^{-\frac{m_c^2}{\alpha(1-\alpha)M^2}},\;
\nonumber \\ & 
\enqa
\begin{equation}
\rho^{(24)}_{\uu}(s)=-\frac{\uu}{6\sqrt{2}}\rho_{pert}^{(22)}(s),\;
\end{equation}
\begin{equation}
\Pi^{(24)}_{\langle G^2\rangle}(M^2)=-\frac{\uu}{6\sqrt{2}}\Pi^{(22)}_{\langle 
G^2\rangle}(M^2)
\end{equation}
\beqa
\Pi^{(24)}_{\langle\bar{u}G u\rangle}(M^2)
& = \frac{5 \langle\bar{q}g_s\sigma\cdot Gq \rangle}{3\cdot 2^6\sqrt{2}\pi^2}
\int_0^1d\alpha\,\frac{m_c^2}{1-\alpha}\,e^{-\frac{m_c^2}{\alpha(1-
\alpha)M^2}},\;
\enqa
\beqa
\rho^{(44)}_{pert}(s)
&=
\frac{3}{2^{12}\pi^6}\int_{\alpha_{min}}^{\alpha_{max}}d\alpha
\int_{\beta_{min}}^{\beta_{max}}d\beta\,\frac{1-\left(\alpha+\beta \right)^2}{
\alpha^3\beta^3}K^4(\alpha,\beta),\;
\nonumber \\ &
\enqa
\beqa
\rho^{(44)}_{\uu}(s)
&=
-\frac{3 m_c \uu}{2^{7}\pi^4}\int_{\alpha_{min}}^{\alpha_{max}}\!d\alpha
\int_{\beta_{min}}^{\beta_{max}}\!d\beta\frac{(1+\alpha+\beta)}{\alpha\beta^2}
\! K^2(\alpha,\beta),\;
\nonumber \\ &
\enqa
\begin{equation}
\rho^{(44)}_{\uu^2}(s)=\frac{m_c^2}{2^4\pi^2}\uu^2\sqrt{1-\frac{4m_c^2}{s}}
,\;
\end{equation}
\beqa
\rho^{(44)}_{\langle G^2\rangle}(s)
& = \frac{\langle g^2G^2\rangle}{2^{11}\pi^6}
\int_{\alpha_{min}}^{\alpha_{max}} d\alpha\int_{\beta_{min}}^{\beta_{max}}d
\beta 
\Bigg[
m_c^2 \frac{(1-(\alpha+\beta)^2)}{\alpha^3}
\nonumber \\ &
-
\frac{1-2\alpha-2\beta}{\alpha\beta^2}K(\alpha,\beta)
\Bigg] K(\alpha,\beta),\;
\nonumber \\ 
\enqa
\beqa
\rho^{(44)}_{\langle\bar{u}Gu\rangle}(s)
& =
-\frac{3m_c}{2^8\pi^4}\langle\bar{u}g\sigma\cdot 
Gu\rangle\int_{\alpha_{min}}^{\alpha_{max}} d\alpha
\Bigg[\frac{2(m_c^2-\alpha(1-\alpha)s)}{1-\alpha}
\nonumber \\ &
-\int_{\beta_{min}}^{\beta_{max}} d\beta\,
\Bigg(1
-2 \frac{\alpha+\beta}{\beta}
\Bigg)\frac{K(\alpha,\beta)}{\beta}\Bigg],\;
\enqa
\beqa
\Pi^{(44)}_{\uu\langle\bar{u}Gu\rangle}(M^2)
& = 
-\frac{m_c^2 \uu\langle\bar{u}g_s\sigma\cdot Gu\rangle}{2^5\pi^2}
\int_0^1 d\alpha
\Bigg[
  \frac{\alpha(1-\alpha)M^2+m_c^2}{\alpha(1-\alpha)M^2}
\nonumber \\ & 
-\frac{1}{1-\alpha}
\Bigg]e^{-\frac{m_c^2}{\alpha(1-\alpha)M^2}}.\;
\label{OPE.last}
\enqa
The integration limits are: 
$$\alpha_{min}=\frac{1-\sqrt{1-
\frac{4m_c^2}{s}}}{2},\qquad\alpha_{max}=\frac{1+\sqrt{1-\frac{4m_c^2}{s}}}{2}$$
$$\beta_{min}=\frac{\alpha}{\alpha\frac{q^2}{m_c^2}-1},\qquad\beta_{max}=1-
\alpha$$
and we define $K(\alpha,\beta)\equiv(\alpha+\beta)m_c^2-\alpha\beta q^2$.

By taking the derivative of Eq.~(\ref{sr})
with respect to $1/M^2$ and dividing the result by Eq.~(\ref{sr}) we can
obtain the mass of $m_X$ without worrying about the value of
the meson-current coupling $\lambda^u$. The expression thus obtained is analised
numerically using the following values for quark masses and QCD condensates
\cite{x3872,narpdg}:
\beqa\label{qcdparam}
&m_c(m_c)=(1.23\pm 0.05)\,\GeV,\nnb\\
&\uu\,-(0.23\pm0.03)^3
\,\GeV^3,\nnb\\
&\lag\bar{u}g\si.Gu\rag=m_0^2\lag\bar{u}u\rag,\nnb\\
&m_0^2=
0.8\,\GeV^2,\nnb\\
&\lag g^2G^2\rag=0.88~\GeV^4.
\enqa

\begin{figure}[ht] 
\vspace{-0.8 cm}
\centerline{
\epsfig{figure=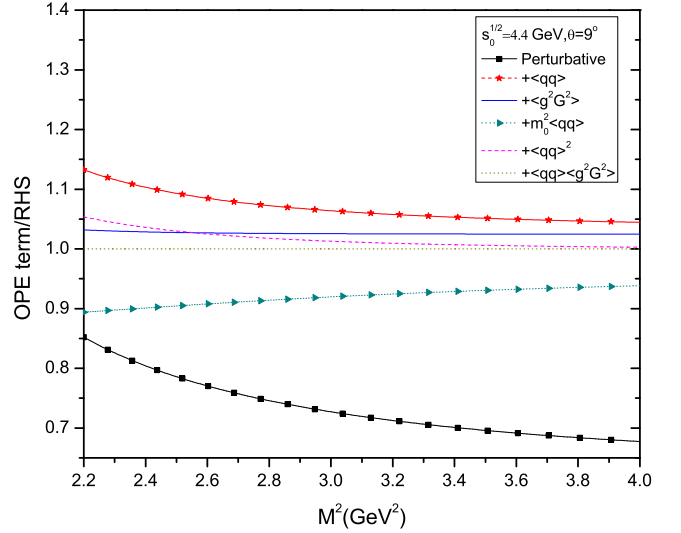}
}
\vspace{-1 cm}
\caption{\label{OPE.conv} Relative contributions of the terms in 
eqs.~(\ref{OPE.first}) 
to (\ref{OPE.last}) grouped by condensate dimensions. We start with the 
perturbative
contribution and each
subsequent line represents the addition of one extra condensate
dimension in the expansion.}
\end{figure} 

In Fig.~\ref{OPE.conv} we show the contributions of the terms in 
Eqs.~(\ref{OPE.first}) 
to (\ref{OPE.last}) grouped by condensate dimensions divided by the RHS of 
Eq.~(\ref{sr}).
We have used $s_{0}^{1/2} = 4.4$ GeV and $\theta = 9^\circ$, but the situation 
does not 
change much for other choices of these parameters.
It is clear that the OPE is converging for values of $M^2 \geq 2.6$ GeV$^2$ and
we will limit our analysis to that region.

\begin{figure}[h] 
\vspace{-0.8 cm}
\centerline{
\epsfig{figure=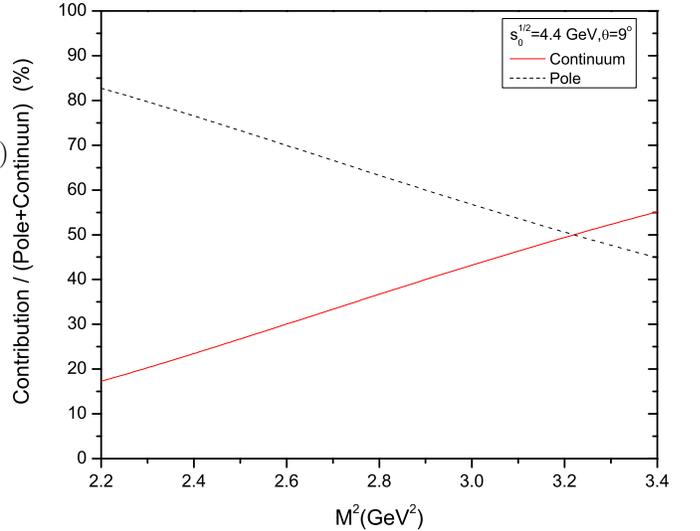}
}
\vspace{-1 cm}
\caption{\label{PVC} The dashed line shows the relative pole contribution (the
pole contribution divided by the total, pole plus continuum,
contribution) and the solid line shows the relative continuum
contribution.}
\end{figure}

The upper limit to the value of $M^2$ comes by imposing that the QCD pole 
contribution
should be bigger than the continuun contribution. The maximum value of $M^2$ 
that satisfies
this condition depends on the value of $s_{0}$, being more restrictive for 
smaller
$s_{0}$. In Fig.~\ref{PVC} we show a comparison between the pole and continuun 
contributions
for the smaller $s_{0}$ we will be considering ($s_{0}^{1/2} = 4.4$) and 
$\theta = 9^\circ$.
The condition obtained from Fig.~\ref{PVC} is $M^2 \leq 3.2$ GeV$^2$, but in 
this case, the dependence on the choice of $\theta$ is very strong. Taking into 
account
the variation of $\theta$ we have determined that, for $5^\circ \leq \theta 
\leq 13^\circ$, the
QCDSR are valid in the following region:
\beq\label{sr.region}
2.6 \GeV^2 \leq M^2 \leq 3.0 \GeV^2
\enq

\begin{figure}[h] 
\vspace{-0.8 cm}
\centerline{
\epsfig{figure=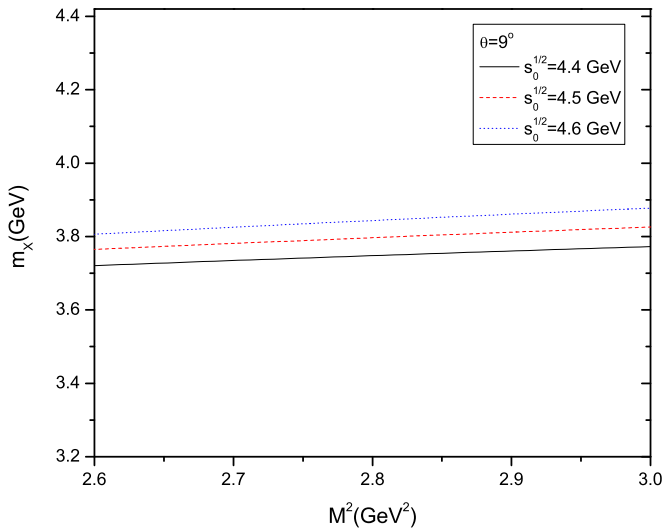}
}
\vspace{-1 cm}
\caption{\label{figmx} The $X$ meson mass as a function of the sum rule 
parameter
($M^2$) in the region of eq.~(\ref{sr.region}) for different values 
of the continuum threshold: $s_0^{1/2} =
4.4$ GeV (solid line), $s_0^{1/2} = 4.5$ GeV (dashed line) and 
$s_0^{1/2} = 4.6$ GeV (dotted line).}
\end{figure}

In Fig.~\ref{figmx}, we show the $X$ meson mass 
in this region. We see that
the results are reasonably stable as a function of $M^2$.
From Fig.~\ref{figmx} we obtain $m_X = (3.80 \pm 0.08) \GeV$ where
the error includes the variation of both $s_0$ and $M^2$. If we also
take into account the variation of $\theta$ in
the region $5^\circ \leq \theta \leq 13^\circ$ we get:
\beq\label{sr.mx}
m_X = (3.77 \pm 0.18) \GeV,
\enq
which is in a good agreement with the experimental value. The value obtained
for the mass grows with the value of the mixing angle $\theta$, but for 
$\theta\geq30^\circ$ it reaches a stable value being completely determined
by the molecular part of the current. 

From Eq.~(\ref{sr}) we can also obtain $\lambda^u$ by fixing $m_X$ equal
to the experimental value ($m_X = 3.87 \GeV$). Using the same region
in $\theta$, $s_0$ and $M^2$ that we have used in the mass analysis we obtain: 
\beq\label{sr.lamb}
\lambda^u = (3.6 \pm 0.9).10^{-3} \GeV^5.
\enq
%

%
%
\section{Decay of the X(3872) and the three point correlator}
%
%
%
As discussed in Sec. I, one of the most intriguing facts about the meson 
$X(3872)$
 is the observation, reported by the BELLE collaboration 
\cite{belleE}, 
that the $X$ decays into $J/\psi\,\pi^+\pi^-\pi^0$, with a strength that
is compatible to that of the  $J/\psi\pi^+\pi^-$ mode, as given by 
Eq~.(\ref{rate}).
This decay suggests an appreciable transition rate to $J/\psi\,\omega$ and 
establishes strong isospin violating effects. It still does not
completely exclude a $c\bar{c}$ interpretation for $X$ since the origin of 
the isospin and G parity non-conservation in Eq.~(\ref{rate}) could be 
of dynamical origin due to $\rho^0-\omega$ mixing \cite{tera}. However,
the observation of the ratio in Eq.~(\ref{rate}) is an important point in favor
of the molecular picture proposed by Swanson \cite{swan1}. In this molecular 
picture the $X(3872)$ is mainly a $D^0\bar{D}^{*0}$ molecule with a small but 
important admixture of $\rho J/\psi$ and $\omega J/\psi$ components. 

It is important to notice that, although a $D^0\bar{D}^{*0}$ molecule is not
an isospin eingenstate, the ratio in Eq.~(\ref{rate}) can not be reproduced by 
a pure $D^0\bar{D}^{*0}$ molecule. This can be seen through the observation that
the decay width for the decay $X\to J/\psi V\to J/\psi F$ where $F=\pi^+\pi^-
(\pi^+\pi^-\pi^0)$ for $V=\rho(\omega)$ is given by \cite{maiani,decayx}
\beqa
&&{d\Gamma\over ds}(X\to J/\psi f)={1\over8\pi m_X^2}|{\cal{M}}|^2B_{V\to F}
\nn\\
&\times&
{\Ga_V m_V\over\pi}{p(s)\over(s-m_V^2)^2+(m_V\Ga_V)^2},
\label{de1}
\enqa
where
\beq
p(s)={\sqrt{\la(m_X^2,m_\psi^2,s)}\over2m_X},
\enq
with $\la(a,b,c)=a^2+b^2+c^2-2ab-2ac-2bc$. The
invariant amplitude squared is given by:
\beq
|{\cal M}|^2=g_{X\psi V}^2f(m_X,m_\psi,s),
\enq
where $g_{X\psi V}$ is the coupling constant in the vertex $XJ/\psi V$ and
\beqa
&&f(m_X,m_\psi,s)={1\over3}\left(4m_X^2-{m_\psi^2+s\over2}
+{(m_X^2-m_\psi^2)^2\over2s}\right.
\nn\\
&+&\left.{(m_X^2-s)^2\over2m_\psi^2}\right){m_X^2-m_\psi^2+s\over2m_X^2}.
\label{m2}
\enqa
Therefore, the ratio in Eq.~(\ref{rate}) is given by:
\beq
{\Gamma(X\to J/\psi\,\pi^+\pi^-\pi^0)\over \Gamma(X\to J/\psi\,\pi^+\pi^-)}
={g_{X\psi\omega}^2m_\omega\Gamma_\omega B_{\omega\to\pi\pi\pi} I_\omega\over
g_{X\psi\rho}^2m_\rho\Gamma_\rho B_{\rho\to\pi\pi} I_\rho},
\label{ratioga}
\enq
where 
\beqa
I_V&=&\int_{(nm_\pi)^2}^{(m_X-m_\psi)^2}ds\bigg(f(m_X,m_\psi,s)
\nn\\
&\times&{p(s)\over(s-m_V^2)^2+(m_V\Ga_V)^2}\bigg).
\enqa

Using $B_{\omega\to\pi\pi\pi}=0.89$, $\Gamma_\omega=8.49\GeV$, $m_\omega=782.6
\MeV$, $B_{\rho\to\pi\pi}=1$, $\Gamma_\rho=149.4\GeV$ and $m_\rho=775.5
\MeV$ we get
\beq
{\Gamma(X\to J/\psi\,\pi^+\pi^-\pi^0)\over \Gamma(X\to J/\psi\,\pi^+\pi^-)}
=0.118\left({g_{X\psi\omega}\over g_{X\psi\rho}}\right)^2.
\label{rationum}
\enq

The couplings, $g_{X\psi V}$, can be evaluated through a QCDSR
calculation for the vertex, $X(3872)J/\psi V$, that centers in
the three-point function given by
\beq
\Pi_{\mu\nu\al}(p,\pli,q)=\int d^4x d^4y ~e^{i\pli.x}~e^{iq.y}
\Pi_{\mu\nu\al}(x,y),
\enq
with
\beq
\Pi_{\mu\nu\al}(x,y)=\lag 0 |T[j_\mu^{\psi}(x)j_{\nu}^{V}(y)
{j_\al^X}^\dagger(0)]|0\rag,
\lb{3po}
\enq
where $p=\pli+q$ and the interpolating fields 
are given by:
\beq
j_{\mu}^{\psi}=\bar{c}_a\gamma_\mu c_a,
\lb{psi}
\enq
\beq
j_{\nu}^{V}={N_V\over2}(\bar{u}_a\gamma_\nu u_a+(-1)^{I_V}\bar{d}_a\gamma_\nu 
d_a),
\lb{vec}
\enq
with $N_\rho=1$, $I_\rho=1$, $N_\omega=1/3$ and $I_\omega=0$.
If $X(3872)$ is a pure $D^0\bar{D}^{*0}$ molecule,
$j_{\al}^{X}$ is given by Eq.~(\ref{curr4}). In this case the only difference in
the OPE side of the sum rule is the factor $N_V$ and, therefore, regardless
the approximations made in the OPE side and the number of terms considered in 
the sum rule one has
\beq
\Pi_{\mu\nu\al}^V(p,\pli,q)=N_V\Pi_{\mu\nu\al}^{OPE}(p,\pli,q).
\enq

To evaluate the phenomenological side of the sum rule we  
insert, in Eq.(\ref{3po}), intermediate states for $X$, $J/\psi$ and $V$. 
We get \cite{decayx}:
\beqa
&&\Pi_{\mu\nu\al}^{(phen)} (p,\pli,q)={i\la_X m_{\psi}f_{\psi}m_Vf_{V}~
g_{X\psi V}
\over(p^2-m_{X}^2)({\pli}^2-m_{\psi}^2)(q^2-m_V^2)}
\nn\\
&\times&\bigg(-\epsilon^{\al
\mu\nu\si}(\pli_\si+q_\si)-\epsilon^{\al\mu\si\ga}{\pli_\si q_\ga q_\nu
\over m_V^2}
\nn\\
&-&\epsilon^{\al\nu\si\ga}{\pli_\si q_\ga\pli_\mu\over m_\psi^2}
\bigg).
\lb{phen}
\enqa

Therefore, for a given structure the sum rule is given by:
\beqa
{i\la_X m_{\psi}f_{\psi}m_Vf_{V}~
g_{X\psi V}
\over(p^2-m_{X}^2)({\pli}^2-m_{\psi}^2)(q^2-m_V^2)}=N_V\Pi^{OPE}(p,\pli,q),
\enqa
from where, considering $m_\rho\simeq m_\omega$ one gets:
\beq
{g_{X\psi \omega}f_\omega\over g_{X\psi \rho} f_\rho}={N_\omega\over N_\rho}=
{1\over3}.
\enq
Using $f_\rho=157\MeV$ and $f_\omega=46\MeV$ we obtain
\beq
{g_{X\psi \omega}\over g_{X\psi \rho} }=1.14,
\enq
and using this result in Eq.~(\ref{rationum}) we finally get
\beq
{\Gamma(X\to J/\psi\,\pi^+\pi^-\pi^0)\over \Gamma(X\to J/\psi\,\pi^+\pi^-)}
\simeq0.15.
\label{ratiofi}
\enq

It is very important to notice that this is a very general result that
does not depend on any approximation in the QCDSR. This result shows that the
admixture of $\rho J/\psi$ and $\omega J/\psi$ components in the molecular
model of ref.\cite{swan1} is indeed very important to reproduce the data
in Eq.~(\ref{rate}). 
It is also important to notice that, in a QCDSR 
calculation of the decay rate $X\to J/\psi V$, the $c\bar{c}$ admixture in the 
$D^0\bar{D}^{*0}$ molecule, as given by Eq.~(\ref{field}), does not solve the 
problem 
of geting the ratio in Eq.(\ref{rate}). This can be seen by using, in 
Eq.~(\ref{3po}), ${j_\al^X}=J_\al^u$, with $J_\al^u$ given by Eq.~(\ref{field}). 
One gets:
\beqa
\Pi_{\mu\nu\al}(x,y)&=&{\uu\over2\sqrt{6}} \cos(\theta)
\Pi^{c\bar{c}}_{\mu\nu\al}(x,y)
\nn\\
&+&\sin(\theta)\Pi^{mol}_{\mu\nu\al}(x,y),
\label{sepi}
\enqa
where
\beq
\Pi^{c\bar{c}}_{\mu\nu\al}(x,y)=\lag 0 |T[j_\mu^{\psi}(x)j_{\nu}^{V}(y)
{j_\al^{'(2)}}^\dagger(0)]|0\rag,
\enq
and
\beq
\Pi^{mol}_{\mu\nu\al}(x,y)=\lag 0 |T[j_\mu^{\psi}(x)j_{\nu}^{V}(y)
{j_\al^{(4u)}}^\dagger(0)]|0\rag,
\enq
with ${j_\al^{'(2)}}$ and ${j_\al^{(4u)}}$ given by Eqs.~(\ref{curr2}) and 
(\ref{curr4}). Using the currents in Eqs.(\ref{vec}) and (\ref{psi}) for 
 the mesons $V$ and $J/\psi$, it is easy to see that
\beqa
\Pi^{c\bar{c}}_{\mu\nu\al}(x,y)&=&{N_V\over2}\Tr\left[\gamma_\mu S^c_{ac}(x)
\gamma_\al\gamma_5S^c_{ca}(-x)\right]\times\nn\\
&\times&\Tr\left[\gamma_\nu S^u_{bb}(0)+(-1)^{I_V}
\gamma_\nu S^d_{bb}(0)\right].
\label{decc}
\enqa

For $V=\rho$ with $I_\rho=1$ the result in Eq.~(\ref{decc}) is obviously zero
due to isospin conservation, in the case that the quark $u$ and $d$ are 
degenerate. However, even for $V=\omega$ ($I_\omega=0$), the result in 
Eq.~(\ref{decc}) is zero because $\Tr\left[\gamma_\mu S^q_{bb}(0)\right]=0$. 
Therefore, in the OPE side, the three-point function is given only by the
molecular part of the current in Eq~(\ref{field}):
\beq
\Pi_{\mu\nu\al}(x,y)=\sin(\theta)\Pi^{mol}_{\mu\nu\al}(x,y),
\enq
that can not reproduce the experimental observation in Eq.~(\ref{rate}), as
demonstrated above.

In the following, to be able to reproduce the data in Eq.(\ref{rate}),
instead of the admixture of $\rho J/\psi$ and $\omega J/\psi$ components to the
$D^0\bar{D}^{*0}$ molecule, as done by Swanson \cite{swan1}, we will consider
a small admixture of $D^+D^{*-}$ and $D^-D^{*+}$ components. In this case, 
instead of Eq.(\ref{field}) we have
\beq
j_{\mu}^X(x)= \cos\alpha J_{\mu}^u(x)+\sin\alpha J_{\mu}^d(x),
\label{4mix}
\enq
with $J_{\mu}^u(x)$ and $J_{\mu}^d(x)$ given by Eq.(\ref{field}).

If we consider the quarks $u$ and $d$ to be degenerate, {\it i.e.}, $m_u=m_d$
and $\uu=\dd$, the change in Eq.(\ref{field}) to Eq.(\ref{4mix}) does not
make any difference in the results in Sec.~III.

By inserting $j_{\mu}^X$, given by Eq.~(\ref{4mix}), in Eq.~(\ref{3po}) and
considering the quarks $u$ and $d$ to be degenerate, one has
\beqa
\Pi_{\mu\nu\al}(p,\pli,q)&=&\sin(\theta){N_V\over2\sqrt{2}}\big(\cos\al
\nn\\
&+&(-1)^{I_V}\sin\al\big)\Pi_{\mu\nu\al}^{OPE}(p,\pli,q),
\label{opemix}
\enqa
with
\beqa
&&\Pi_{\mu\nu\al}^{OPE}(p,\pli,q)=\int~d^4u\int~{d^4k\over(2\pi)^4}\bigg(\Tr\big
[
\gamma_\mu S^c_{a'c}(k)\gamma_5\times
\nn\\
&\times&S^q_{ab'}(-y)\gamma_\nu S^q_{b'b}(y)\gamma_\al
S^c_{ba'}(k-p')\big]+
\nn\\
&-&\Tr\big[\gamma_\mu S^c_{a'c}(k)\gamma_\al S^q_{ab'}(-y)\gamma_\nu S^q_{b'b}(y)
\gamma_5 S^c_{ba'}(k-p')\big]\bigg).
\enqa

In the phenomenological side, considering the definition of $\lambda^u$ in 
Eq.(\ref{lau}) and the definition of the current in (\ref{4mix}), we
can define
\beq
\lambda_X=\cos\al\lambda^u+\sin\al\lambda^d=(\cos\al+\sin\al)\lambda^q,
\label{laX}
\enq
where $\lambda^q$ was evaluated in Sec.~III, and is given in 
Eq.~(\ref{sr.lamb}). 
Using Eq.(\ref{laX}) in Eq.(\ref{phen}), the phenomenological side of the sum 
rule is now given by:
\beqa
&&\Pi_{\mu\nu\al}^{(phen)} (p,\pli,q)={i(\cos\al+\sin\al)\lambda^q m_{\psi}
f_{\psi}m_Vf_{V}~g_{X\psi V}
\over(p^2-m_{X}^2)({\pli}^2-m_{\psi}^2)(q^2-m_V^2)}
\nn\\
&\times&\bigg(-\epsilon^{\al
\mu\nu\si}(\pli_\si+q_\si)-\epsilon^{\al\mu\si\ga}{\pli_\si q_\ga q_\nu
\over m_V^2}
\nn\\
&-&\epsilon^{\al\nu\si\ga}{\pli_\si q_\ga\pli_\mu\over m_\psi^2}
\bigg).
\lb{phenmix}
\enqa

From Eqs.~(\ref{opemix}) and (\ref{phenmix}) we get the following relation
between the coupling constants:
\beq
{g_{X\psi \omega}f_\omega\over g_{X\psi \rho} f_\rho}={N_\omega\big(\cos\al+\sin
\al\big)\over N_\rho\big(\cos\al-\sin\al\big)}.
\enq
 Using the previous result in Eq.~(\ref{rationum}) and the numerical values for
$f_\omega$ and $f_\rho$ we have
\beq
{\Gamma(X\to J/\psi\,\pi^+\pi^-\pi^0)\over \Gamma(X\to J/\psi\,\pi^+\pi^-)}
\simeq0.15\left({\cos\al+\sin\al\over\cos\al-\sin\al}\right)^2.
\label{ratioal}
\enq
This is exactly the same relation obtained in refs.~\cite{maiani,decayx}, that
determines $\al\sim20^0$ for reproducing the experimental result in 
Eq.(\ref{rate}). A similar relation was obtained in ref.~\cite{oset} where
the decay of the $X$ into two and three pions goes through a $D$ $D^*$ loop. 

With this mixing angle $\al$ defined, we can now evaluate the decay rate itself,
for any one of the decays: $X\to J/\psi\rho$ or $X\to J/\psi\omega$, since they 
will be the same. Therefore, we choose to work with $X\to J/\psi\omega$
since the combination $\cos\al+\sin\al$ appears in both sides of the sum rule
and the result for $g_{X\psi\omega}$ is independent of $\al$.

\begin{widetext}
\begin{center} 
\begin{figure}[h] 
\epsfig{figure=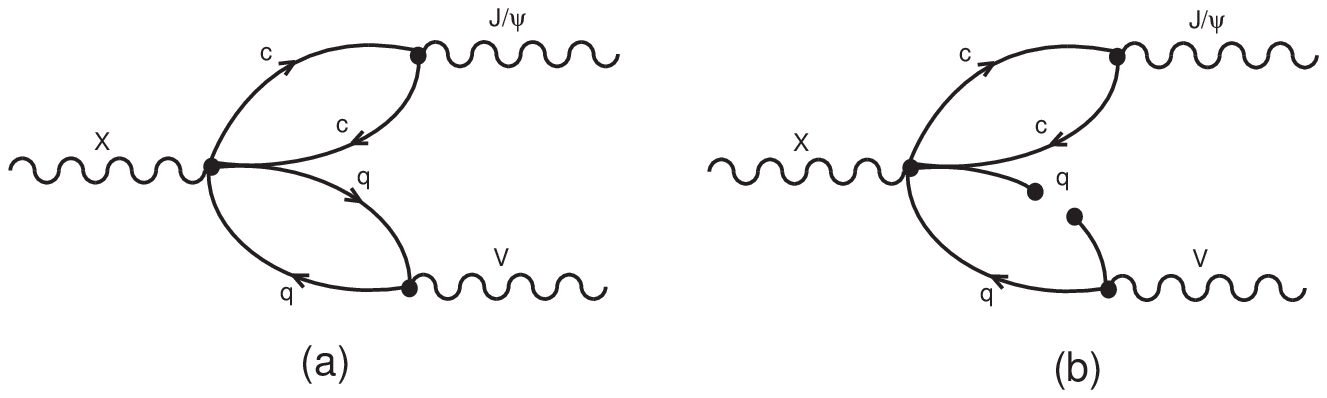,height=30mm}
\epsfig{figure=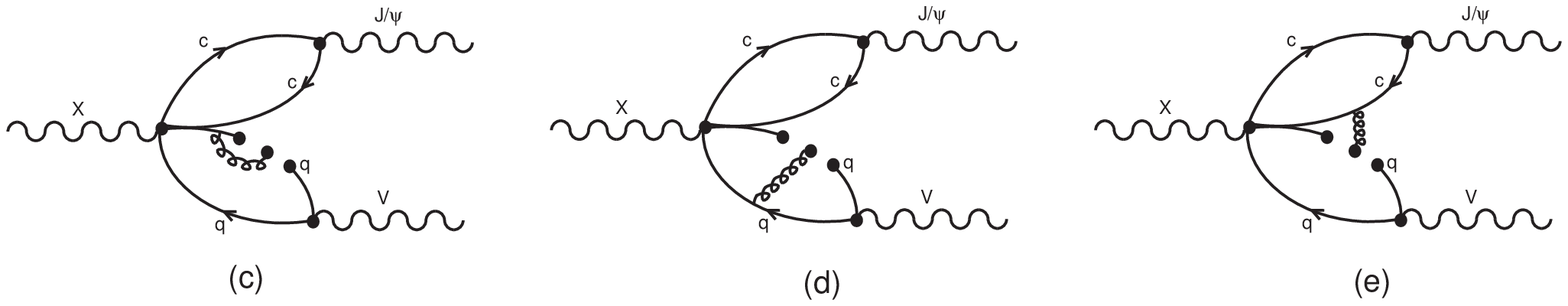,height=30mm}
\caption{\label{3pdiags} Diagrams which contribute to the OPE side of the sum 
rule.}
\end{figure}   
\end{center}   
\end{widetext}

In the OPE side we consider condensates up to dimension five , as shown in 
Fig.~\ref{3pdiags}. 
Taking the limit $p^2={\pli}^2=-P^2$ and doing a
single Borel transform to $P^2\rightarrow M^2$, we get  
in the structure $\epsilon^{\al\nu\si\ga}\pli_\si q_\ga\pli_\mu$
(the same considered in ref.\cite{decayx}) ($Q^2=-q^2$):
\beqa
C(Q^2)\left(e^{-m_\psi^2/M^2}-e^{-m_X^2/M^2}\right)+B~e^{-s_0/M^2}=
\nn\\
(Q^2+m_\omega^2)\Pi^{(OPE)}(M^2,Q^2),
\label{3sr}
\enqa
where
\beqa
&&\Pi^{(OPE)}(M^2,Q^2)={\lag\bar{q}q\rag\over6\sqrt{2}\pi^2Q^2}\bigg[\bigg({m_0^
2
\over3Q^2}+
\nn\\
&-&1\bigg)\int_{4m_c^2}^{u_0}du~e^{-u/M^2}~\sqrt{1-4m_c^2/u}
\left({1\over2}+{m_c^2
\over u}\right)+
\nonumber\\
&-&{m_0^2\over16}\int_0^1 d\al{1+3\al\over\al}~e^{-m_c^2\over
\al(1-\al)M^2}\bigg].
\label{est1}
\enqa

In Eq.~(\ref{3sr})
\beq
C(Q^2)={6\over\sin(\theta)}m_\omega f_\omega{f_\psi\lambda^q\over m_\psi
(m_X^2-m_\psi^2)}g_{X\psi\omega}(Q^2),
\label{CXV}
\enq
and $B$ gives the contribution of the pole-continuum transitions 
\cite{decayx,dsdpi,io2}. $s_0$ and $u_0$ 
are the continuum thresholds for $X$ and $J/\psi$ respectively. Notice that
in Eq.(\ref{est1}) we have introduced the form factor $g_{X\psi\omega}(Q^2)$.
This is because the meson $\omega$ is off-shell in the vertex $XJ/\psi\omega$.

\begin{figure}[h] 
\vspace{-0.8 cm}
\centerline{
\epsfig{figure=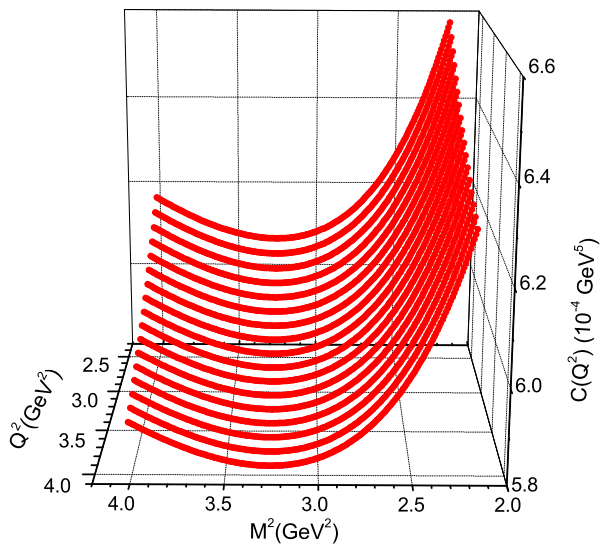}
}
\vspace{-1 cm}
\caption{\label{fig.3D} Values of $C(Q^2)$ obtained
by varying both $Q^2$ and $M^2$ in Eq.~(\ref{3sr}).
}
\end{figure}

If we parametrize $C(Q^2)$ as a monopole:
\beq
C(Q^2)= {c_1\over Q^2+c_2},
\label{parC}
\enq
we can fit the left hand side of Eq.~(\ref{3sr}) as a function of $Q^2$ and 
$M^2$ to the QCDSR results in the right hand side, 
obtaining $c_1$, $c_2$ and $B$. 
In Fig.~\ref{fig.3D} we show the points obtained if we isolate $C(Q^2)$ in
Eq.~(\ref{3sr}) and vary both $Q^2$ and $M^2$. 
The function $C(Q^2)$ (and
consequently $g_{X\psi\omega}(Q^2)$) should not
depend on $M^2$, so we limit our fit region to 
$3.0 \GeV^2 \leq M^2 \leq 3.5 \GeV^2$ where $C(Q^2)$
is clearly stable in $M^2$ for all values of
$Q^2$.

We do the fitting for $s_0^{1/2} = 4.4 \GeV$ as
the results do not depend much on this parameter,
the results are shown bellow:
\beqa
&c_1 = 2.5 \times 10^{-2} \GeV^7, 
\nnb \\
&c_2 = 38 \GeV^2, 
\nnb \\ 
&B = 2.9 \times 10^{-4} \GeV^5.
\label{fitF}
\enqa

\begin{figure}[h] 
\vspace{-0.8 cm}
\centerline{
\epsfig{figure=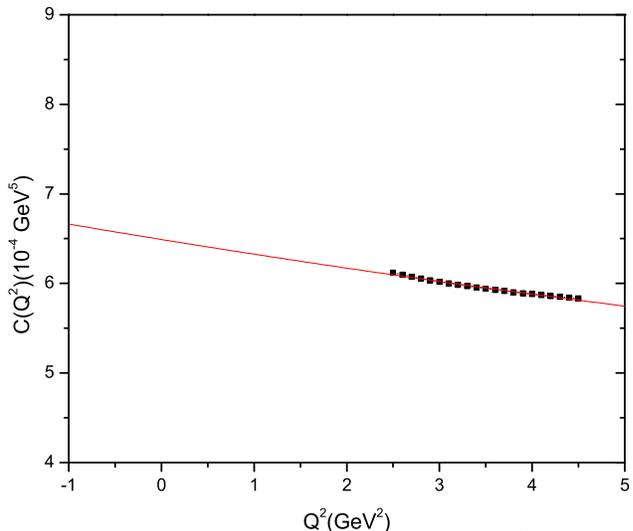}
}
\vspace{-1 cm}
\caption{\label{fig.fitF} Momentum dependence of  $C(Q^2)$ for
$s_0^{1/2} = 4.4 \GeV$. The solid line gives the 
parametrization of the QCDSR results (dots) through Eq.~(\ref{parC}) 
and (\ref{fitF}).}
\end{figure}

In Fig.~\ref{fig.fitF} we can see that the $Q^2$ dependence of $C(Q^2)$ is
well reproduced by the chosen parametrization in the 
interval $2.5\leq Q^2\leq4.5~\GeV^2$, where the QCDSR is valid.

The form factor $g_{X\psi\omega}(Q^2)$  can then be easily obtained by using
Eqs.~(\ref{CXV}) and (\ref{parC}).  Since the coupling constant is
defined as the value of the form factor at the meson pole: $Q^2=-m_\omega^2$,
to determine the coupling constant we have to extrapolate $g_{X\psi\omega}(Q^2)$
to a $Q^2$ region where the sum rules are no longer valid (since the QCDSR 
results are valid in the deep Euclidian region). Using $m_{\psi} = 3.1 \GeV$,
$m_{X} =3.87 \GeV$, $f_{\psi} = 0.405 \GeV $, $\lambda^u = 3.6 \times 10^3 
\GeV^5$ 
from Eq.~(\ref{sr.lamb}) and varying $\theta$ in the range $5^\circ \leq \theta 
\leq 13^\circ$, we get:
\beq
g_{X \psi \omega} = g_{X\psi\omega}(-m_\omega^2) = 5.4 \pm 2.4
\label{sr.coupl}
\enq 

The decay width is given by:
\beq
\Gamma\left(X\rightarrow J/\psi \pi^+ \pi^- \pi^0 \right) = 
g_{X \psi \omega}^2 {m_{\omega} \Gamma_{\omega} \over 8 \pi^2 m^2_X} 
B_{\omega \rightarrow \pi \pi \pi } I_{\omega},
\enq
which, together with Eq.~(\ref{sr.coupl}) gives us:
\beq
\Gamma\left(X\rightarrow J/\psi \pi^+ \pi^- \pi^0 \right) = (9.3 \pm 6.9)~\MeV.
\label{regama}
\enq

The result in Eq.~(\ref{regama}) is in complete agreement with the experimental
upper limit. It is important to notice that the width grows with the
mixing angle $\theta$, as can be seen from Eq.~(\ref{CXV}), while the mass
grows with $\theta$. Therefore, there is only a small range for the values
of this angle that can provide simultaneously good agreement with the
experimental values of the mass and the decay width, and this range is 
$5^\circ \leq \theta \leq 13^\circ$. This means that the $X(3872)$ is basically 
a $c\bar{c}$ state with a small, but fundamental, admixture of molecular 
$DD^*$ states. By molecular states we mean an admixture between  
$D^0\bar{D}^{*0},~\bar{D}^0{D}^{*0}$ and $D^+D^{*-},~D^-D^{*+}$ states, as 
given by Eq.~(\ref{4mix}). 

%
%
\section{Conclusions}
%
%
We have presented a QCDSR analysis of the two-point and three-point
functions of the $X(3872)$ meson, by considering a mixed charmonium-molecular
current.  We find that the sum rules results in Eqs.~(\ref{sr.mx}) and
(\ref{regama}) are compatible with experimental data. These results were 
obtained by considering the mixing angle in Eq.~(\ref{field}) in the range
$5^\circ \leq \theta \leq 13^\circ$. 

We have also studied the mixing between the $D^0\bar{D}^{*0},~\bar{D}^0{D}^{*0}$
and $D^+D^{*-},~D^-D^{*+}$ states by imposing the ratio in Eq.~(\ref{rate}).
In accordance with the findings in ref.~\cite{maiani} we found that the mixing
angle in Eq.~(\ref{4mix}) is $\alpha\sim20^0$.

With the knowledge of these two mixing angles we conclude that the $X(3872)$ is 
basically a $c\bar{c}$ state ($\sim$97\%) with a small, but fundamental, 
admixture of molecular $D^0\bar{D}^{*0},~\bar{D}^0{D}^{*0}$ ($\sim$88\%) and 
$D^+D^{*-},~D^-D^{*+}$ ($\sim$12\%) states. 

This small molecular component
could, in principle, be a consequence of neglecting the two-hadron reducible 
contribution in the phenomenological side.
However, as argued in section III, we expect the 2HR contribution to be small
and the results to hold even if we had taken it into consideration.


\begin{thebibliography}{999}

\bibitem{belle1} S.-K. Choi  {\it et al.} [Belle Collaboration],
Phys. Rev. Lett. {\bf 91}, 262001 (2003).

\bibitem{Xexpts} 
 V.~M.~Abazov {\it et al.}  [D0 Collaboration],
  Phys.\ Rev.\ Lett.\  {\bf 93}, 162002 (2004);
D. Acosta {\it et al.} [CDF Collaboration], Phys. Rev. Lett. 
{\bf 93}, 072001 (2004);
  B.~Aubert {\it et al.}  [BaBar Collaboration],
  Phys.\ Rev.\ D {\bf 71}, 071103 (2005).

\bibitem{belleE}
  K.~Abe {\it et al.} [Belle Collaboration], hep-ex/0505037,
hep-ex/0505038.

\bibitem{babar2}  B. Aubert {\it et al.}  [BaBar Collaboration],
  Phys.\ Rev.\ D {\bf 74}, 071101 (2006).

\bibitem{cdf2} D. Abulencia {\it et al.} [CDF Collaboration], Phys. Rev. 
Lett. {\bf 96}, 102002 (2006).

\bibitem{cdf3} D. Abulencia {\it et al.} [CDF Collaboration], Phys. Rev. 
Lett. {\bf 98}, 132002 (2007).

\bibitem{bg} T. Barnes and S. Godfrey, Phys. Rev. D {\bf 69}, 054008 (2004). 

\bibitem{cleo} C. Cawfield {\it et al.}  [CLEO Collaboration],
  Phys.\ Rev.\ Lett. {\bf 98}, 092002 (2007).

\bibitem{close} F.E. Close and P.R. Page, Phys.\ Lett. B {\bf578}, 119 
(2004).

\bibitem{swanson} E.S. Swanson,
  Phys.\ Rept.\  {\bf 429}, 243 (2006).

\bibitem{maiani} L. Maiani, F. Piccinini, A.D. Polosa, V. Riquer, 
Phys. Rev. D {\bf 71}, 014028 (2005).

\bibitem{polosa} A.D. Polosa, arXiv:hep-ph/0609137.

\bibitem{belleB0} K.~Abe {\it et al.} [Belle Collaboration], arXiv:0809.1224.

\bibitem{babarB0}  B.~Aubert {\it et al.}  [BaBar Collaboration],
arXiv:0803.2838.

\bibitem{x3872} R.D. Matheus, S. Narison, M. Nielsen and J.-M. Richard,
 Phys. Rev. D {\bf 75}, 014005 (2007).

\bibitem{lnw} S.H. Lee, M. Nielsen and U. Wiedner, arXiv:0803.1168.


\bibitem{oka24}
  J.~Sugiyama, T.~Nakamura, N.~Ishii, T.~Nishikawa and M.~Oka, 
  Phys.\ Rev.\  D {\bf 76}, 114010 (2007)
  [arXiv:0707.2533 [hep-ph]].


\bibitem{suzuki}  M.~Suzuki,
  Phys.\ Rev.\  D {\bf 72}, 114013 (2005).

\bibitem{swan1} E.S. Swanson, Phys.\ Lett. B {\bf588}, 189 (2004); Phys.\ Lett.
B {\bf598}, 197 (2004).

\bibitem{li1}  R. Li and K.-T. Chao,  Phys.\ Rev.\  D {\bf 79}, 114020 
(2009).

\bibitem{li2}  B.-Q. Li, C. Meng and K.-T. Chao,  arXiv:0904.4068.

\bibitem{liuliu} Y.-R. Liu, X. Liu, W.-Z. Deng and S.-L. Zhu,
Eur. Phys. J. C {\bf56}, 63 (2008).

\bibitem{dong} Y. Dong, A. Faessler, T. Gutsche and V.E. Lyubovitskij,
Phys. Rev. D {\bf77}, 094013 (2008).

\bibitem{stancu} Fl. Stancu, arXiv:0809.0408.

\bibitem{babar09} B.~Aubert {\it et al.}  [BaBar Collaboration], Phys. Rev.
Lett. {\bf102}, 132001 (2009).

\bibitem{svz} M.A. Shifman, A.I. and Vainshtein and V.I. Zakharov,
Nucl. Phys. B {\bf 147}, 385 (1979).

\bibitem{rry} L.J. Reinders, H. Rubinstein and S. Yazaki, Phys. Rept.
{\bf 127}, 1 (1985).

\bibitem{SNB} For a review and references to original works, see
e.g., S.
Narison, {\it QCD as a theory of hadrons,
Cambridge Monogr. Part. Phys. Nucl. Phys. Cosmol.} {\bf 17}, 1 (2002)
[hep-h/0205006]; {\it QCD
spectral sum rules ,  World Sci. Lect. Notes Phys.} {\bf 26}, 1 (1989);
{ Acta Phys. Pol.} {\bf B26}, 687 (1995); { Riv. Nuov. Cim.} {\bf 10N2}, 1
(1987); { Phys. Rept.} {\bf 84}, 263 (1982).

\bibitem{2hr} Y. Kondo, O. Morimatsu, T. Nishikawa,  Phys. Lett. {\bf B611}, 
93 (2005).

\bibitem{2hr2} S.H. Lee, H. Kim, Y. Kwon, Phys. Lett. {\bf B609}, 252 (2005).


\bibitem{narpdg} S. Narison, Phys. Lett. {\bf B466}, 345 (1999);
S. Narison,  Phys. Lett. {\bf B361}, 121 (1995);
S. Narison, Phys. Lett. {\bf B387}, 162 (1996); S. Narison,  Phys.  Lett.
{\bf B624}, 223 (2005).

\bibitem{tera} K. Terasaki, arXiv:0904.3368.

\bibitem{decayx} F.S. Navarra, M. Nielsen, Phys. Lett. {\bf B639}, 272 (2006).

\bibitem{oset} D. Gamermann, E. Oset, arXiv:0905.0402.

\bibitem{dsdpi} M. Nielsen, Phys. Lett. {\bf B634}, 35 (2006).

\bibitem{io2} B.~L. Ioffe and A.V. Smilga, Nucl.\ Phys.\ {\bf B232}, 109
 (1984).
\end{thebibliography}
\end{document}